\documentclass[prb,letterpaper,twocolumn,aps, 10pt]{revtex4-1}

\usepackage[english]{babel}	
\usepackage{times}

\usepackage[colorlinks=true, a4paper=true, pdfstartview=FitV,linkcolor=blue, citecolor=blue, urlcolor=blue]{hyperref}

\usepackage{amsmath}		
\usepackage{amssymb}		
\usepackage{amsmath}	
\usepackage{bbm}			

\usepackage{graphicx}
\DeclareGraphicsExtensions{.pdf}

\newcommand{\resp}{\emph{resp.~}}
\newcommand{\Figref}[1]{Fig.~\ref{#1}}
\newcommand{\bea}{\begin{eqnarray}}
\newcommand{\eea}{\end{eqnarray}}
\renewcommand{\Im}{\mathrm{Im}}

\newcommand{\Align}[2]{\begin{align}\label{#1}#2\end{align}}
\newcommand{\Eqref}[1]{\eqref{#1}}

\begin{document}
\title{Semi-Meissner state and non-pairwise intervortex interactions in type-1.5 superconductors}
\author{
Johan Carlstrom${}^{1,2}$, Julien Garaud${}^{2,1}$, Egor Babaev${}^{2,1}$}
\affiliation{
${}^1$ Department of Theoretical Physics The Royal Institute of Technology Stockholm SE-10691 Sweden\\
${}^2$ Department of Physics University of Massachusetts Amherst MA 01003 USA 
}

\begin{abstract}
We demonstrate existence of  {\it non-pairwise} interaction forces 
between   vortices in  multicomponent and layered superconducting 
systems.
That is, in contrast to most common models, the interactions 
in a group of such vortices is not a universal superposition 
of Coulomb or Yukawa forces.
Next we consider the properties of vortex clusters in Semi-Meissner state 
of type-1.5 two-component superconductors.
 We show that under certain condition 
non-pairwise forces can contribute to formation of 
very complex vortex states in type-1.5 regimes.
\end{abstract}
\maketitle
\section{Introduction}
The crucial importance of  topological excitations in the physics of 
 superconductivity made quantum vortex solutions in the Ginzburg-Landau (GL)  models  perhaps the most studied examples of   topological solitons  
(defined as  localized lumps of energy characterized by a topological invariant) \cite{manton,degenes}.
 These  well studied vortex solutions are frequently used as generic 
testing objects for High Energy Physics and Cosmological models \cite{manton}. 
In that broader context, 
especially  spectacular theoretical works attempted on numerous occasions to identify 
topological solitons in the Skyrme and Faddeev models  with  particles \cite{manton}.
 There the particle-solitons are  lumps of energy 
which enjoy a topological protection against radiating their energy. 
Superconducting vortices, in spite being known to form a variety of ``aggregate" vortex 
states: vortex liquids, glasses etc still do not show nearly as complex 
ground states as e.g. the Skyrme or 
Faddeev models \cite{manton}
because of the
smaller diversity of known intervortex 
interaction potentials. The recent works on multicomponent superconductors aimed at 
realizing  more complex bound multi-vortex states  in the so-called 
``type-1.5 regime". In that regime, due to existence of two components, 
thermodynamically stable vortex solutions 
were found that exhibit strongly non-monotonic interaction potentials between two vortices, 
with short-range repulsive and long-rage attractive parts \cite{bs1} and thus form 
vortex clusters in low magnetic fields. 
After the recent experimental publications \cite{m1}
this physics  received increased attention (see e.g. \cite{m2}).

In this work we investigate the structure of multi-vortex bound states in   two-component 
superconductors beyond the validity of the linearized theory and find  
{\it non-pairwise multi-body interactions} which are especially pronounced at short 
intervortex separations.  We show that in certain cases the non-pairwise 
forces are strong enough to led  to extremely rich multivortex states.
We also propose how these vortex states can be experimentally realized in layered superconducting structures.

We consider a Ginzburg-Landau model of a two-component superconductor 
which appears in 
various physical systems ranging from multiband and layered superconductors to projected 
states of metallic hydrogen and models of neutron stars interior \cite{bs2}.
\begin{eqnarray}
\label{ind_energy}
\mathcal{F}&=&  
\frac{1}{2}\sum_{i=1,2}\Biggl[|(\nabla+ ie{\bf A}) \psi_i  |^2
+ (2\alpha_i+\beta_i|\psi_i|^2)|\psi_i|^2\Biggr]\nonumber \\
&+&\frac{1}{2}(\nabla \times {\bf A})^2 -\eta|\psi_1|| \psi_2|\cos(\theta_2-\theta_1) 
\end{eqnarray}
Here $e$ is the electric charge, $\psi_{1,2}=|\psi_{1,2}|e^{i\theta_{1,2}}$ represent 
the superconducting  components coupled by the gauge field ${\bf A}$ and the Josephson 
coupling $\eta$.  For a microscopic derivation of such model for two-band superconductors see \cite{gurevich}.
It was demonstrated recently in the framework or a self-consistent microscopic theory
that this ``minimalistic" two-band model describes qualitatively well intervortex interaction
in multiband superconductor in a rather wide range of temperatures  \cite{silaev}.
We choose the units where $\hbar=c=1$. 

\begin{figure}
\includegraphics[width=80mm]{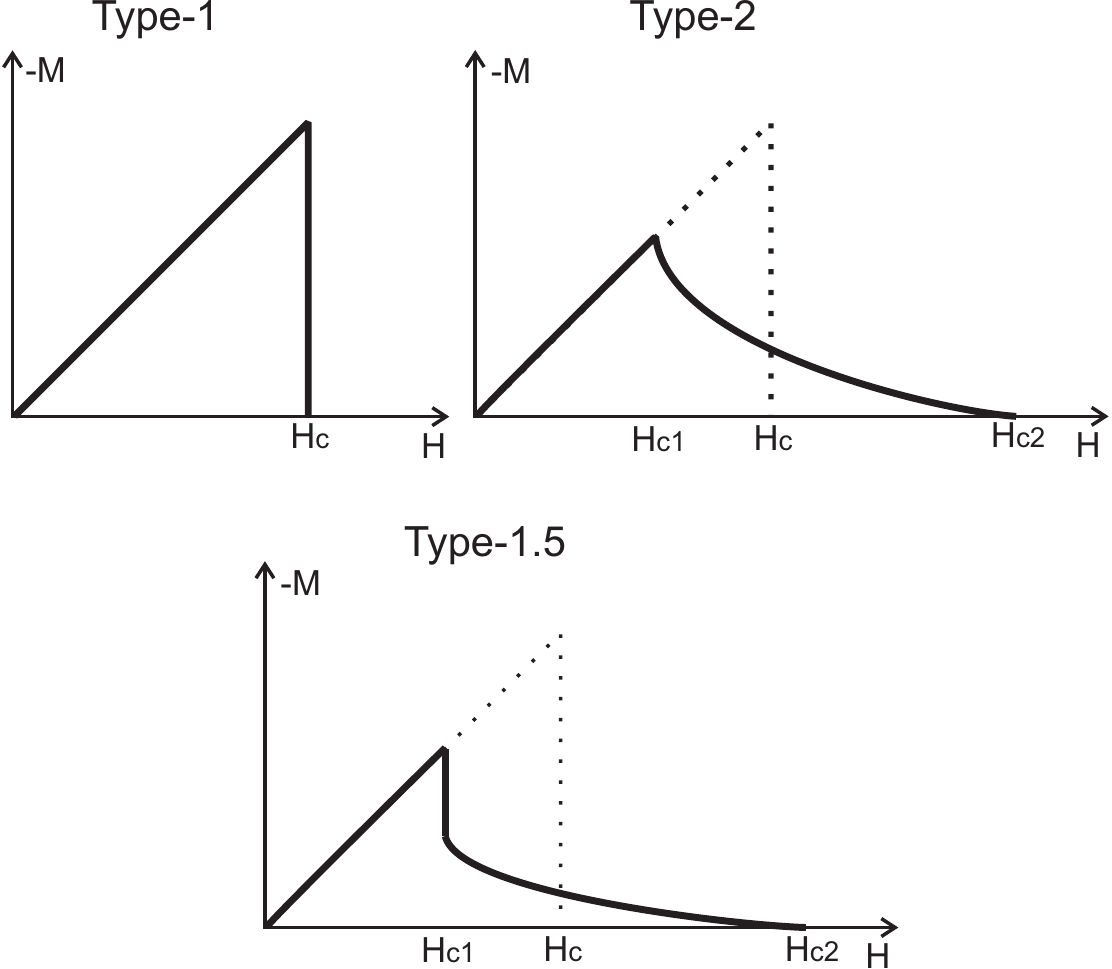}
\caption{A schematic picture of
magnetization curves of type-I, type-II and type-1.5 superconductors.
Type-1.5 superconductor has a first order phase transition in low magnetic fields.
In low magnetic field it can support a phase separation into vortex droplets 
and Meissner domains (the semi-Meissner state).
}
\label{magnetization}
\end{figure}

The key feature of the model  is that one-flux quantum 
vortices induced by magnetic field are ``composite". That is, they have a core around 
which both phases wind by $2\pi$, i.e. 
$\Delta \theta_{1,2}=\oint_{\mathcal{C}}\nabla \theta_{1,2}\mathrm{d}\ell= 2\pi\,$  
($\mathcal{C}\,$ being a closed path around the vortex core). One-quanta vortex can 
thus be viewed here as a bound state of two ``fractional" vortices \cite{frac,Nature,Nature2}. 
The model exhibits type-1.5 superconductivity 
when the magnetic field penetration length scale $\lambda$ is smaller than one of the characteristic length 
scales  of the density variation $\xi_{1,2}$ and also the conditions for short range repulsion and 
thermodynamical stability are satisfied \cite{bs1}. In the type-1.5 regime two vortices 
with similar circulation have interaction that is attractive at long range (driven by 
density-density interaction) but repulsive at short range (due to current-current and 
magnetic interaction) \cite{bs1}. This results in a formation of a  ``semi-Meissner" 
state in low magnetic fields where vortex clusters coexist with two-component Meissner state. 
The magnetization curves of a type-1.5 superconductor are schematically shown on Fig. \ref{magnetization}.

\begin{figure}
 \hbox to \linewidth{ \hss
 \includegraphics[width=\linewidth]{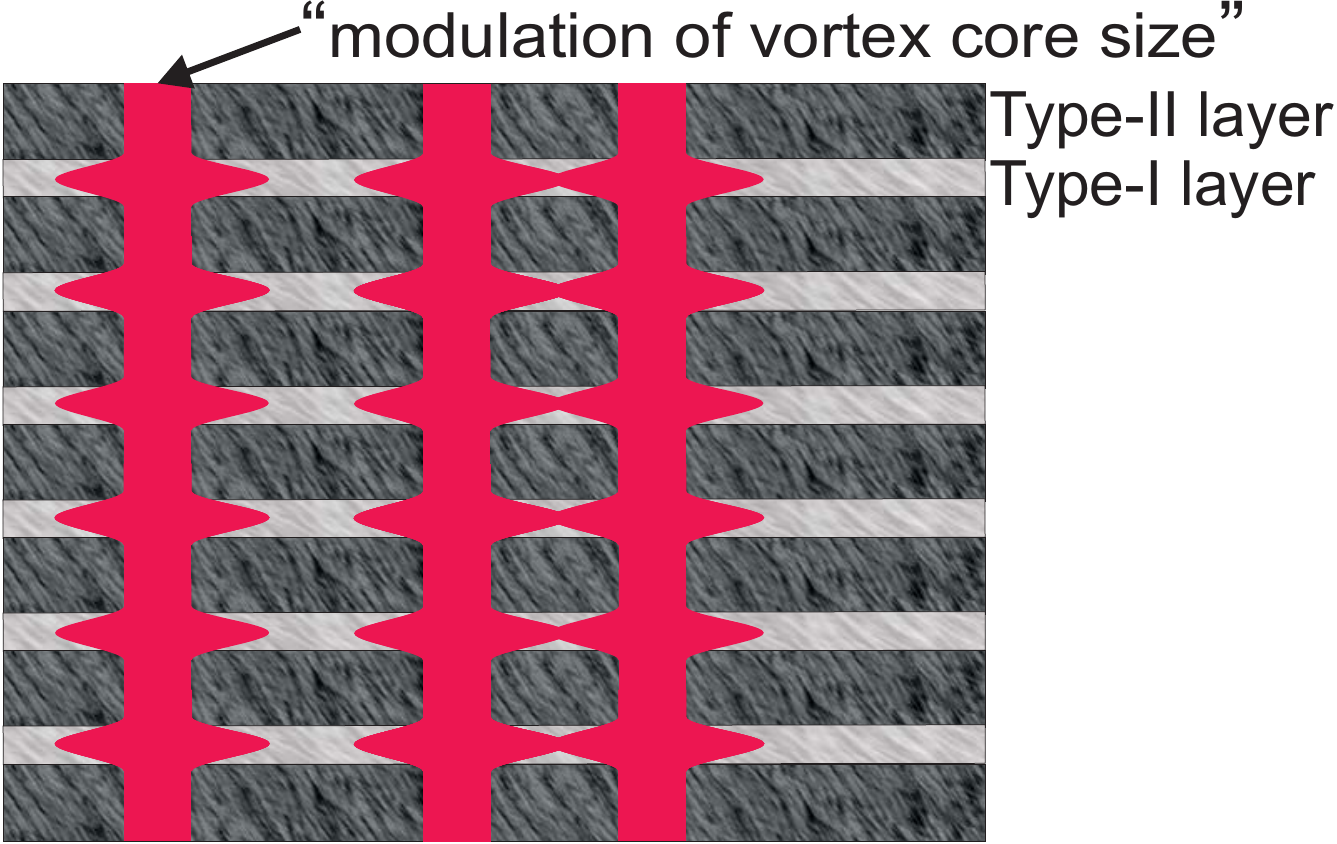}
 \hss}
\caption{A schematic picture of a  collection of Josephson-coupled layers of type-I and type-II superconductors.
Vortices induced by magnetic field are kept in stacks by interlayer Josephson and electromagnetic coupling. If in the type-I layers the cores
will extend beyond the average magnetic field penetration length, these 
extended cores should cause attractive interaction between these vortex lines. The system can be used to model the type-1.5 behavior.
 }
\label{layers}
\end{figure} 


\begin{figure*}
 \includegraphics[width=\linewidth]{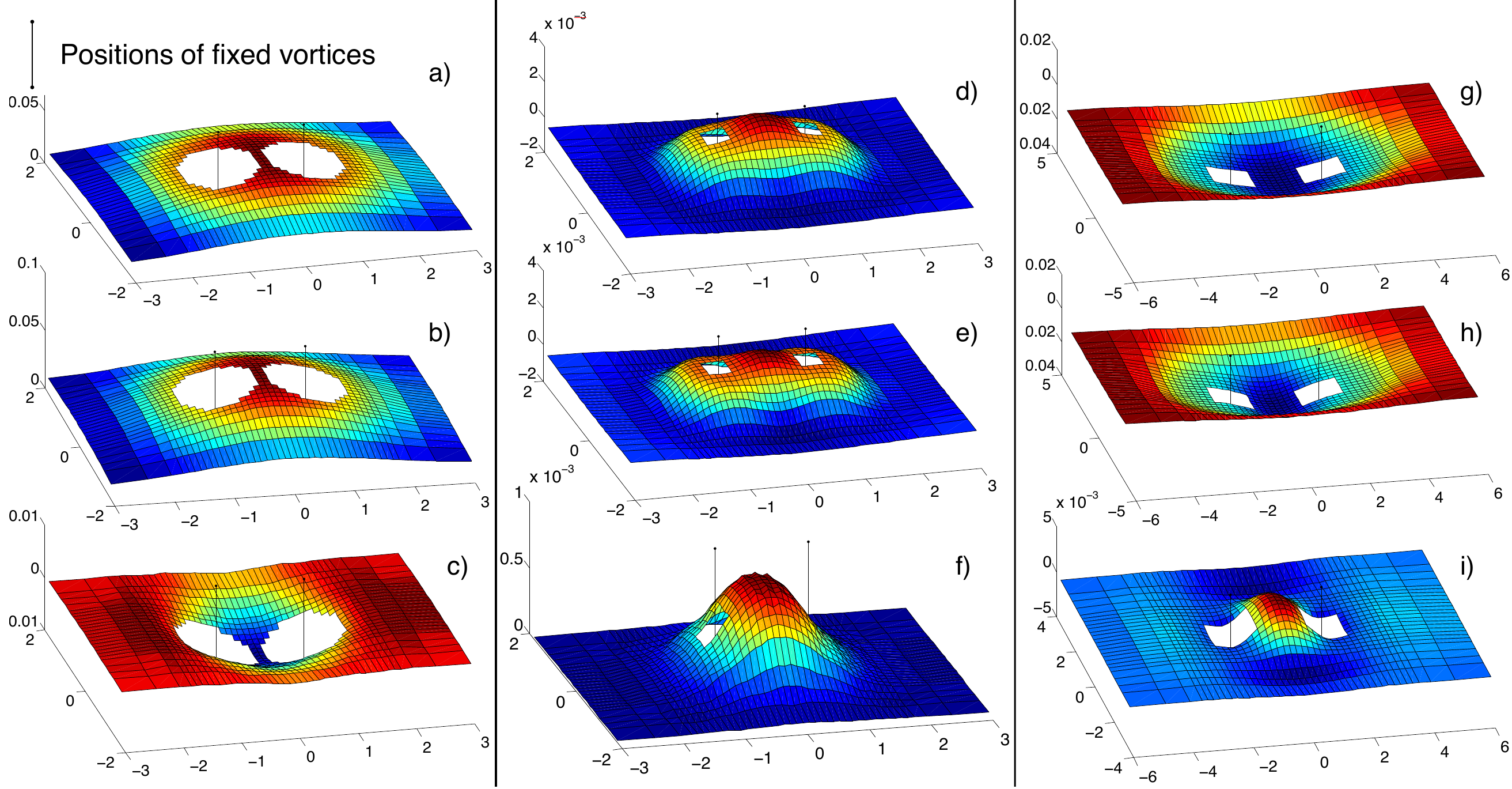}
 \caption{Interaction energy between a single vortex and a fixed vortex pair 
(position of the fixed vortices are shown by two black lines).
GL parameters are : $(\alpha_1,\beta_1)=(-1.0,1.0)$ and $e=1.3$; 
$(\alpha_2,\beta_2)=(3.0,0.5)$, $\eta=3$ for panels $\bf a\,$, $\bf b\,$, $\bf c\,$; 
$(\alpha_2,\beta_2)=(3.0,0.5)$, $\eta=7$ for $\bf d\,$, $\bf e\,$, $\bf f\,$ panels and 
$(\alpha_2,\beta_2)=(-0.0625,0.25)$, $\eta=0.5$ for panels $\bf g\,$, $\bf h\,$, $\bf i\,$. 
Top row ($\bf a\,$, $\bf d\,$, $\bf g\,$) gives the total interaction energy, while second 
one ($\bf b\,$, $\bf e\,$, $\bf h\,$) gives the sum of pairwise interactions. The third row 
($\bf c\,$, $\bf f\,$, $\bf i\,$) displays the difference between these energies, which demonstrates
the existence of strong nontrivial, non-pairwise, (three body) interactions.
The regimes shown in the left, middle and right columns are  type-II, type-1.5 and type-I correspondingly. }
\label{pan5}
\end{figure*}

We consider  below two types of systems: (i) two-band and layered superconducting systems
with weak and strong Josephson coupling and (ii) the
systems where there is no Josephson coupling. The latter situation occurs  in the context of the theories 
of liquid metallic hydrogen, deuterium and their mixtures \cite{Nature,Nature2}, 
and physics of neutron star interiors, where the two fields 
represent electronic, protonic and $\Sigma^-$ hyperon Cooper pair condensates \cite{jones} (for literature overview see \cite{bs2}). 
Because one cannot convert, say, electrons to protons or deuterons,
the intercomponent Josephson coupling in this case is forbidden
 on symmetry grounds but the condensates are coupled by the vector potential.
 Experimentally the type-1.5-like physics  can 
be also realized in  a system of interlaced Josephson-coupled 
type-I/type-II multilayers shown schematically in \Figref{layers}.

The usual framework within which vortex matter in superconductors and superfluids 
is usually discussed relies on the assumption that interactions in a system of vortices
is a superposition of pairwise forces.
Indeed the most usual analysis of interaction between well separated topological 
solitons involves linearization. By nature of this approximation, the interaction 
in a system of multiple vortices is a superposition of two-body forces. 
Similarly, 
for example the description  of phase transitions in type-II superconductors in terms of vortex loop 
proliferation is based on London approximation \cite{peskin}. 
In this approximation the 
fluctuations of densities are neglected and  the intervortex interaction is then pairwise.
Non-linear effects which lead to nonpairwise contributions in the intervortex interaction are, in general, less discussed.
For certain vortex configurations in a single-component Ginzburg-Landau model,
 nonlinear effects  were studied  recently  in \cite{peeters}.

Here we demonstrate the importance 
of  complicated non-pairwise forces between superconducting vortices arising 
in multicomponent systems.
  It has particularly important consequences for 
vortex clusters in the type-1.5 regime (though also relevant for type-II two band
superconductors). 
 
\begin{figure*}[!htb]
  \hbox to \linewidth{ \hss
 \includegraphics[width=1.0\linewidth]{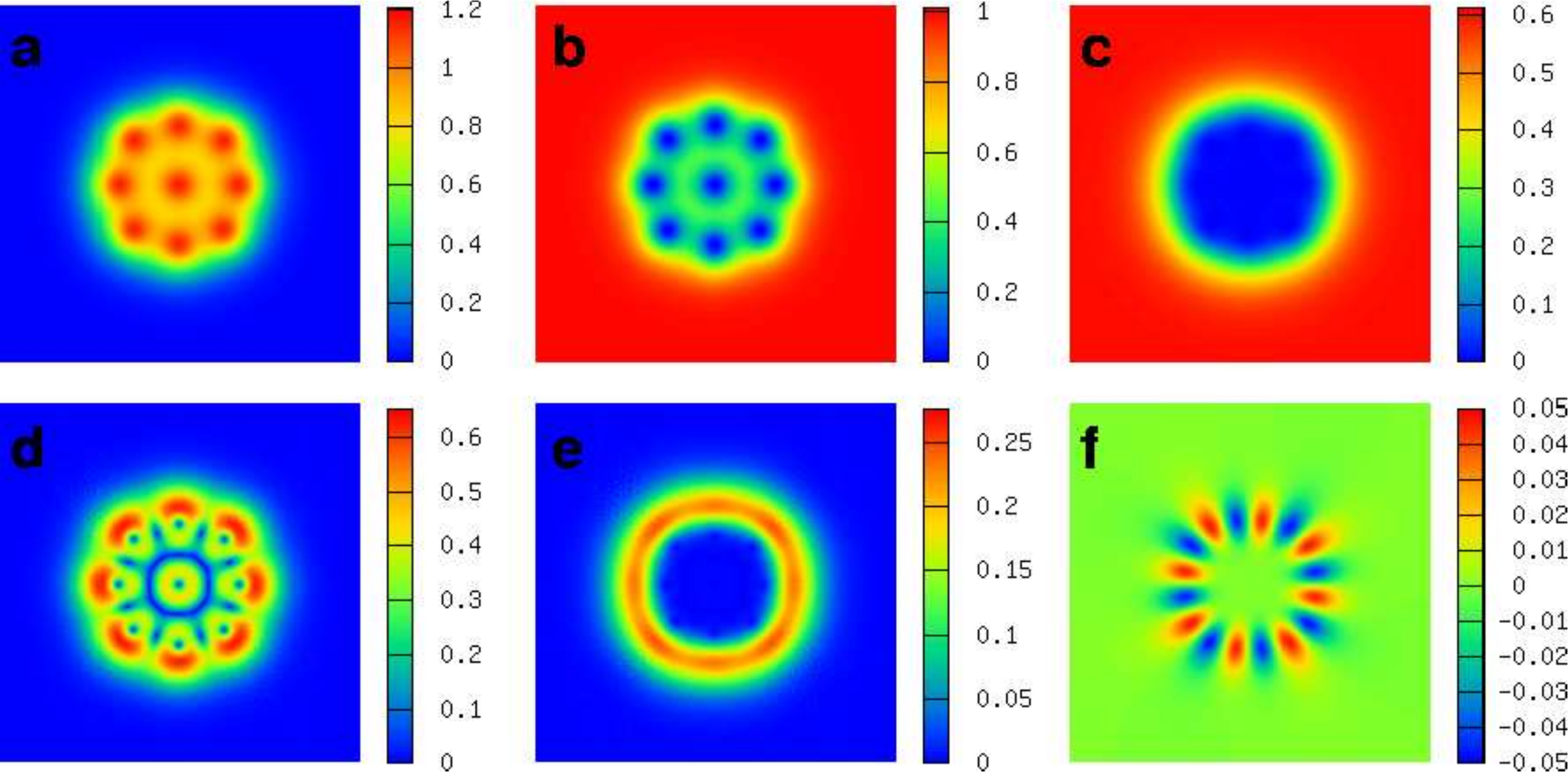}
 \hss}
\caption{
The ground state of a $N_v=9$ flux quanta configuration in type-1.5 $U(1)\times U(1)$
superconductor (i.e. $\eta=0$). The parameters of the potential 
are $(\alpha_1,\beta_1)=(-1.00, 1.00)$ and $(\alpha_2,\beta_2)=(-0.60, 1.00)$, while the electric 
charge is $e=1.48$.  
The physical quantities displayed here are $\bf a$ the magnetic flux density, 
$\bf b$ (\resp $\bf c$) is the density of the first (\resp second) condensate $|\psi_{1,2}|^2$. 
$\bf d$ (\resp $\bf e$) shows the norm of the supercurrent in the  first (\resp second) component.
The panel 
$\bf f$ is $\Im(\psi_1^*\psi_2)$ which is nonzero when there 
appears a difference between the phases of $\psi_1^*$ and $\psi_2$. 
The second component has a type-I like behavior: its density is depleted in the vortex
cluster and its current is mostly concentrated on the boundary of the cluster.
}
\label{2A-1}
\end{figure*}

\section{Three-body intervortex forces}
First, let us present a highly accurate numerical study of a three body vortex
problem. The interaction between vortices was investigated as follows (for
a detailed description see apendix \ref{Numerics-fd}): First a 
vortex pair is fixed in the center of the system. A third vortex is then inserted, 
and the energy is minimized with respect to all the degrees of freedom, except the 
positions of the centers of the vortex cores in the first component.
{ 
The procedure is then repeated for  different positions of the third vortex. 
The resulting interaction energy is shown on \Figref{pan5}
which shows pronounced violation
of the superposition principle for intervortex forces in this system.
To minimize the effects of discretization   the calculation was performed 
on a ultra-high resolution grid of up to $\propto 10^7$ points, with lattice 
spacing $h\sim\xi_1/100$ (where $\xi_1$ is the coherence length of the dominant component 
in the limit of no coupling to the second component). We used 4-16 h on a 8-core cluster 
node to relax each data point in the interaction potential. We 
choose  a ``minimally invasive" procedure of pinning a vortex
on a numerical grid, which gives most accurate
long- and medium- range forces but, on the other hand, does not 
work for the (irrelevant for this study) very short intervortex distances  (for details see Appendix \ref{Numerics-fd}).
Consequently  in \Figref{pan5} no data is given for too closely placed vortices.

\begin{figure*}[!htb]
  \hbox to \linewidth{ \hss
 \includegraphics[width=1.0\linewidth]{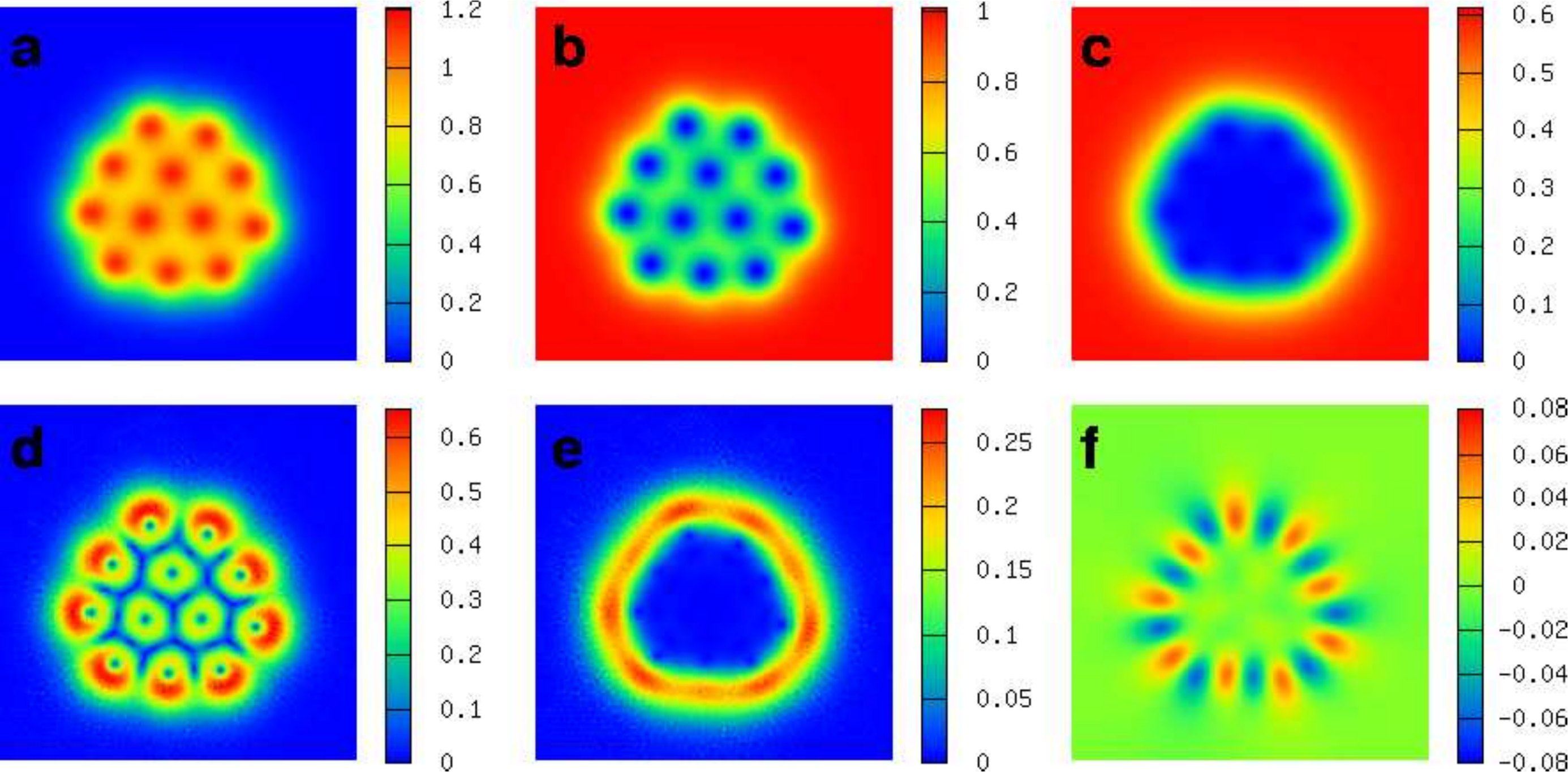}
 \hss}
\caption{
The ground state of a $N_v=12$ flux quanta configuration. The  parameter set is the same as in 
\Figref{2A-1}.
Here the global $8$-folded discrete symmetry of the cluster has been broken down to  
the $3$-folded   symmetry. There is a competition between {type-I-like} (normal circular cluster
with a boundary current) and { type-II-like} tendencies (vortex lattice).
}
\label{2A-2}
\end{figure*}

In all the type-1.5 regimes which we considered we found diverse and pronounced to various degrees
non-pairwise interactions. 

\section{Vortex clusters in a Semi-Meissner state and non-pairwise interactions.}
Let us now investigate how the presence of non-pairwise 
interactions along with non-monotonic two-body forces affects the magnetic response.
Below we report highly accurate numerical solutions for N-vortex bound states in 
several   regimes  (for technical details see Appendix \ref{Numerics-fem}).
Animations showing the evolution of the system, during the numerical
energy minimization, from the various  initial configurations to
the vortex clusters   are also available as supplementary 
material \cite{julien}.

The figures (\Figref{2A-1}--\Figref{fig-new5}) show various bound states of multiple flux 
quanta in $U(1)\times U(1)$ as well as in Josephson-coupled $U(1)$ models. Consider 
first the case where  $\alpha_{1,2}<0$. There appear very 
interesting geometrical properties of the vortex ground states 
(shown on \Figref{2A-1},\Figref{2A-2}). 
One can clearly see that with growing number of vortices, the local vortex structure 
strongly depends on the number of vortices in a cluster. The striking feature which is 
to various degrees is manifest on all the figures   is the coexistence and 
competition between {type-II-like} behavior of the first condensates which attempts to 
form a regular vortex lattice and type-I-like behavior of the second condensate. Namely 
the second condensate mimics the formation of a single large normal domain. Also like in 
a genuine type-I superconductor, this component has supercurrent density 
which is predominantly concentrated on 
the   boundary of the domain. Clearly it energetically prefers a formation of a circular boundary. These competing tendencies in the 
type-1.5 system result  in neither hexagonal nor circular boundary. The next visually striking 
effect is that the vortex solutions are very different inside the vortex cluster and on the 
cluster boundary. This shows up especially clearly in the  current density.

\begin{figure*}[!htb]
  \hbox to \linewidth{ \hss
 \includegraphics[width=1.0\linewidth]{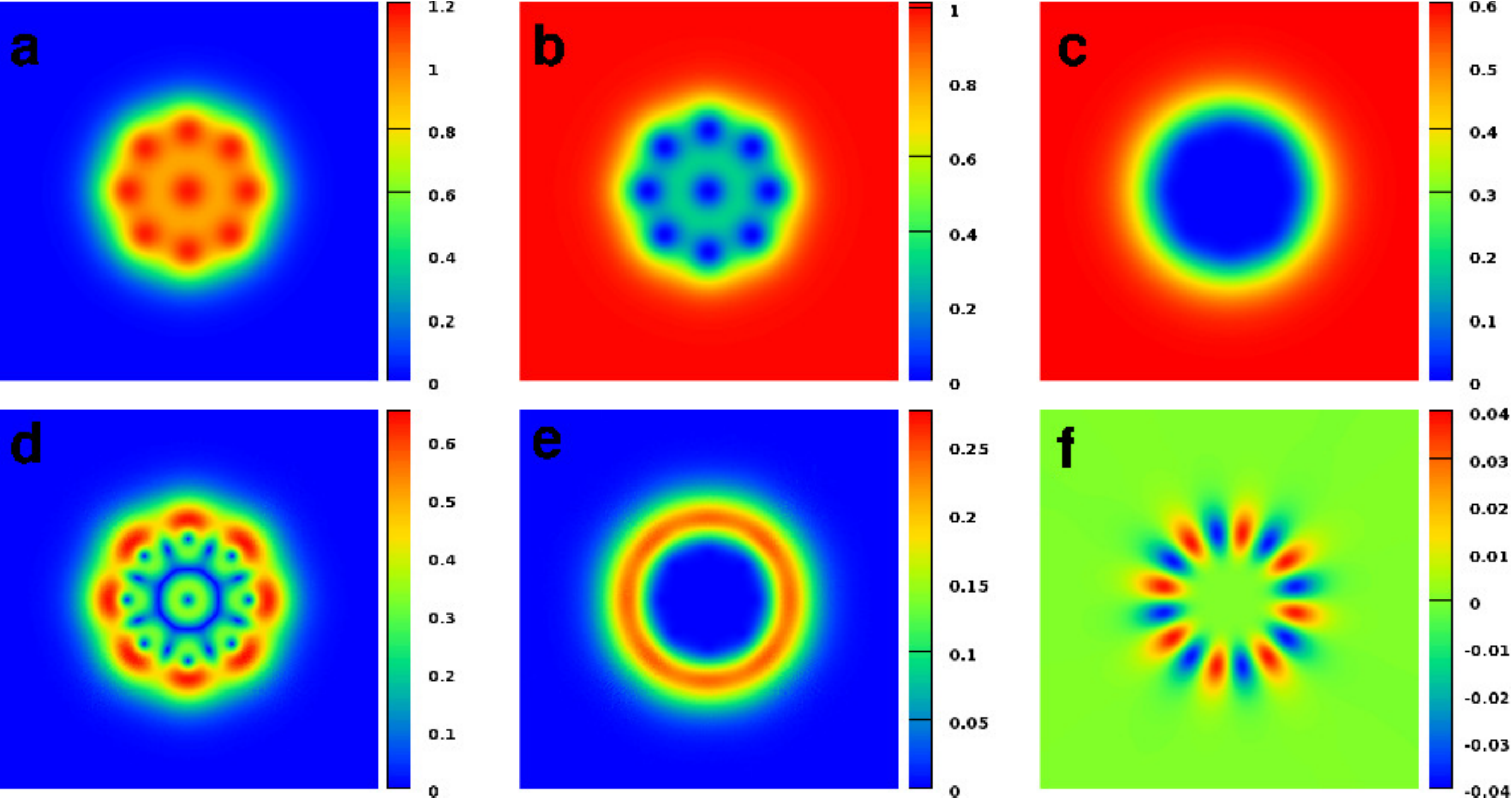}
 \hss}
\caption{
  Ground state of $9$ vortices in a $U(1)\times U(1)$ superconductor with two active bands $(\alpha_2,\beta_2)=(-0.6,1)$ (without  
interband coupling). The parameter set here is as in \Figref{2A-1} except $e=1.55$ 
which places the system closer to the transition with the type-I regime. 
This is manifest in the more circular boundary of the cluster compared to \Figref{2A-1}.
}
\label{fig-new1}
\end{figure*}

\begin{figure*}[!htb]
  \hbox to \linewidth{ \hss
   \includegraphics[width=1.0\linewidth]{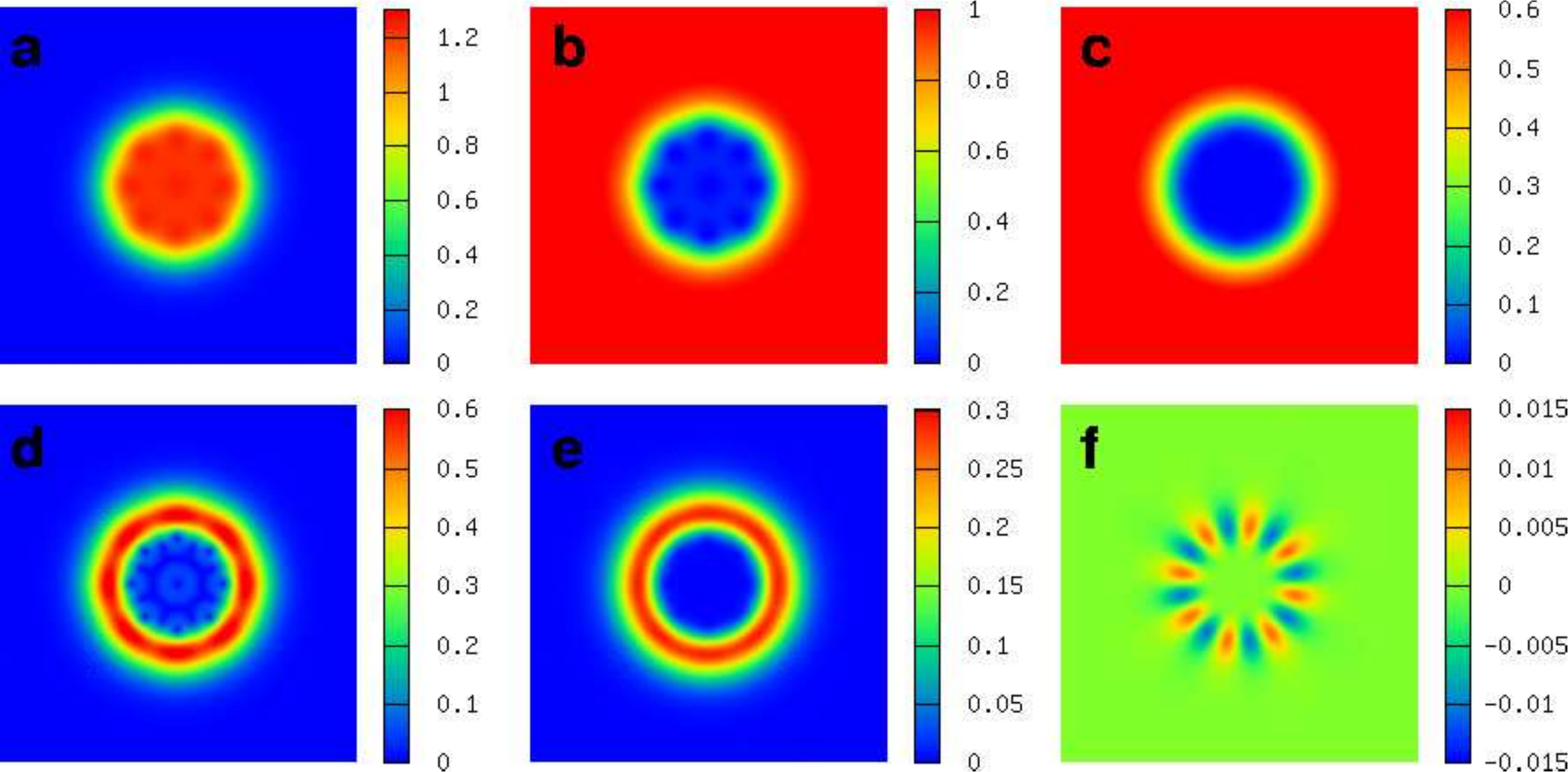}
 \hss}
\caption{
Ground state of an $N_v=9$ vortex configuration with the parameter set given by \Figref{2A-1}, 
but with $e=1.59$ and added  Josephson coupling  $\eta=0.1$.
Although the Josephson term introduced an energy penalty for phase difference, it has little 
effect on the vortex cluster boundary. At the boundary the competing magnetic and density-density interactions win 
over phase-locking terms and generates phase difference gradients. 
Also in this case the system is close to type-I regime: i.e. most 
of the current is concentrated on the boundary of the cluster.  
}
\label{2AWJ-1}
\end{figure*}

\begin{figure*}[!htb]
  \hbox to \linewidth{ \hss
 \includegraphics[width=1.0\linewidth]{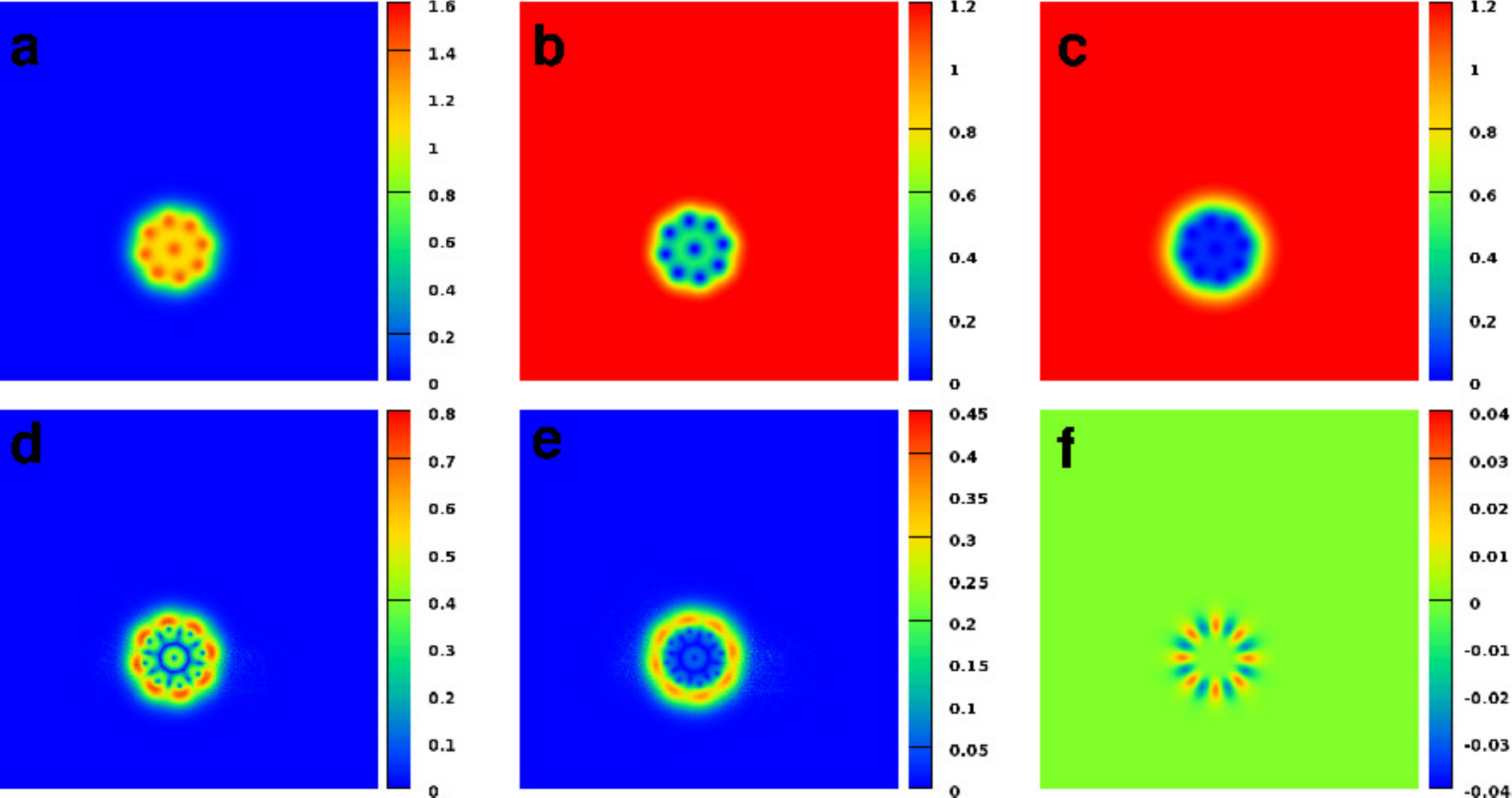}
 \hss}
\caption{
Ground state of $9$ vortices in a superconductor with two active bands. Parameters of the interacting potential 
are $(\alpha_1,\beta_1)=(-1.00, 1.00)$, $(\alpha_2,\beta_2)=(-0.0625, 0.25)$ while the interband coupling is $\eta=0.5$
which is substantially 
larger than on \Figref{2AWJ-1}. The electric charge, parameterizing the penetration depth of the magnetic field, is $e=1.30$
so that the well in the nonmonotonic interacting potential is very small. In this case
there is visible admixure of the current in second component in vortices inside the cluster.
}
\label{fig-new2}
\end{figure*}

\begin{figure*}[!htb]
  \hbox to \linewidth{ \hss
 \includegraphics[width=1.0\linewidth]{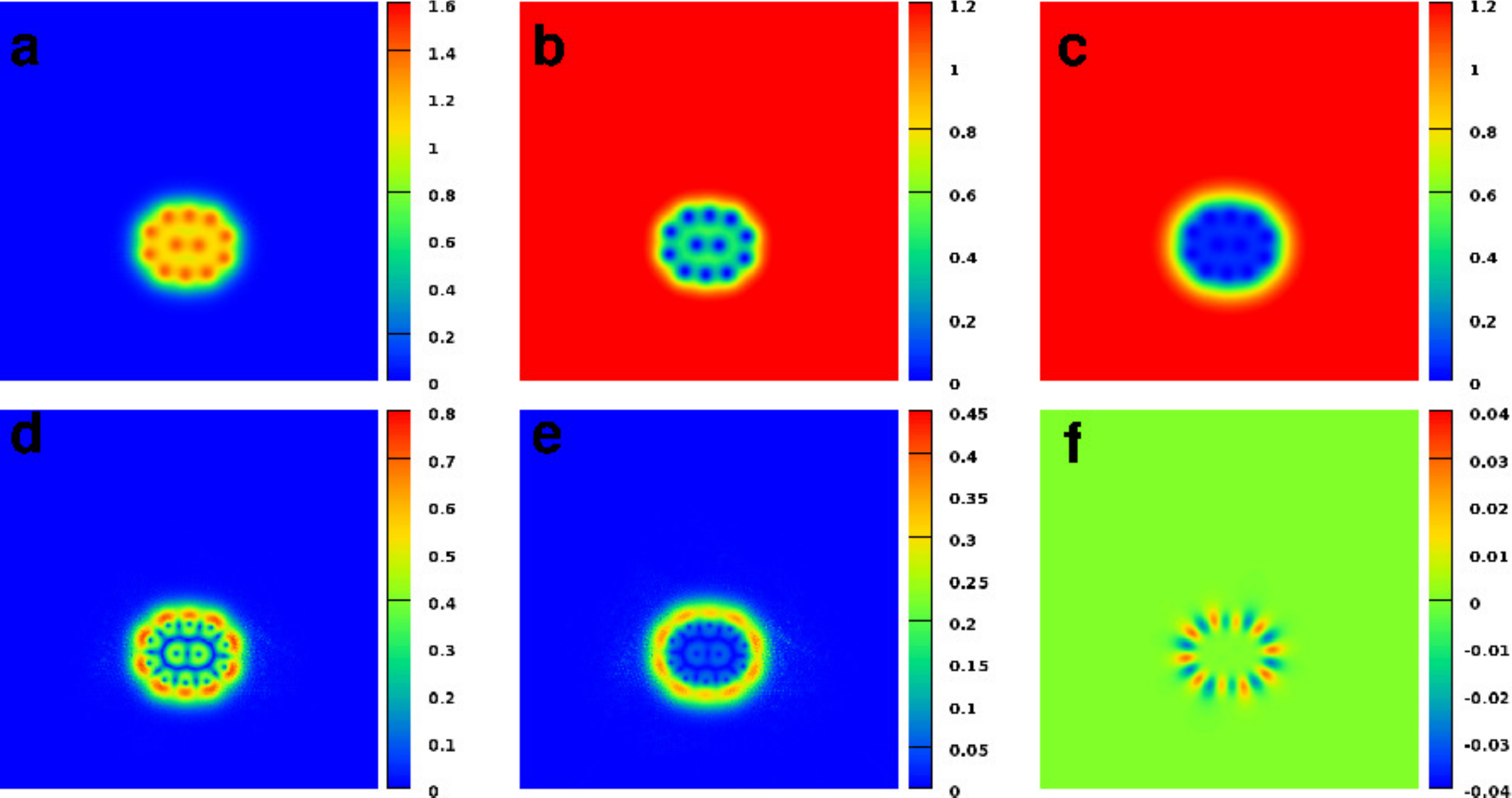}
 \hss}
\caption{
Ground state of $N_v=12$ vortices, for a superconductor with broken $U(1)\times U(1)$ symmetry, 
with the parameters as in \Figref{fig-new2}. 
The system compromises between both type-I-like and type-II-like tendencies in the  different superconducting components.
}
\label{fig-new3}
\end{figure*}

\begin{figure*}[!htb]
  \hbox to \linewidth{ \hss
 \includegraphics[width=1.0\linewidth]{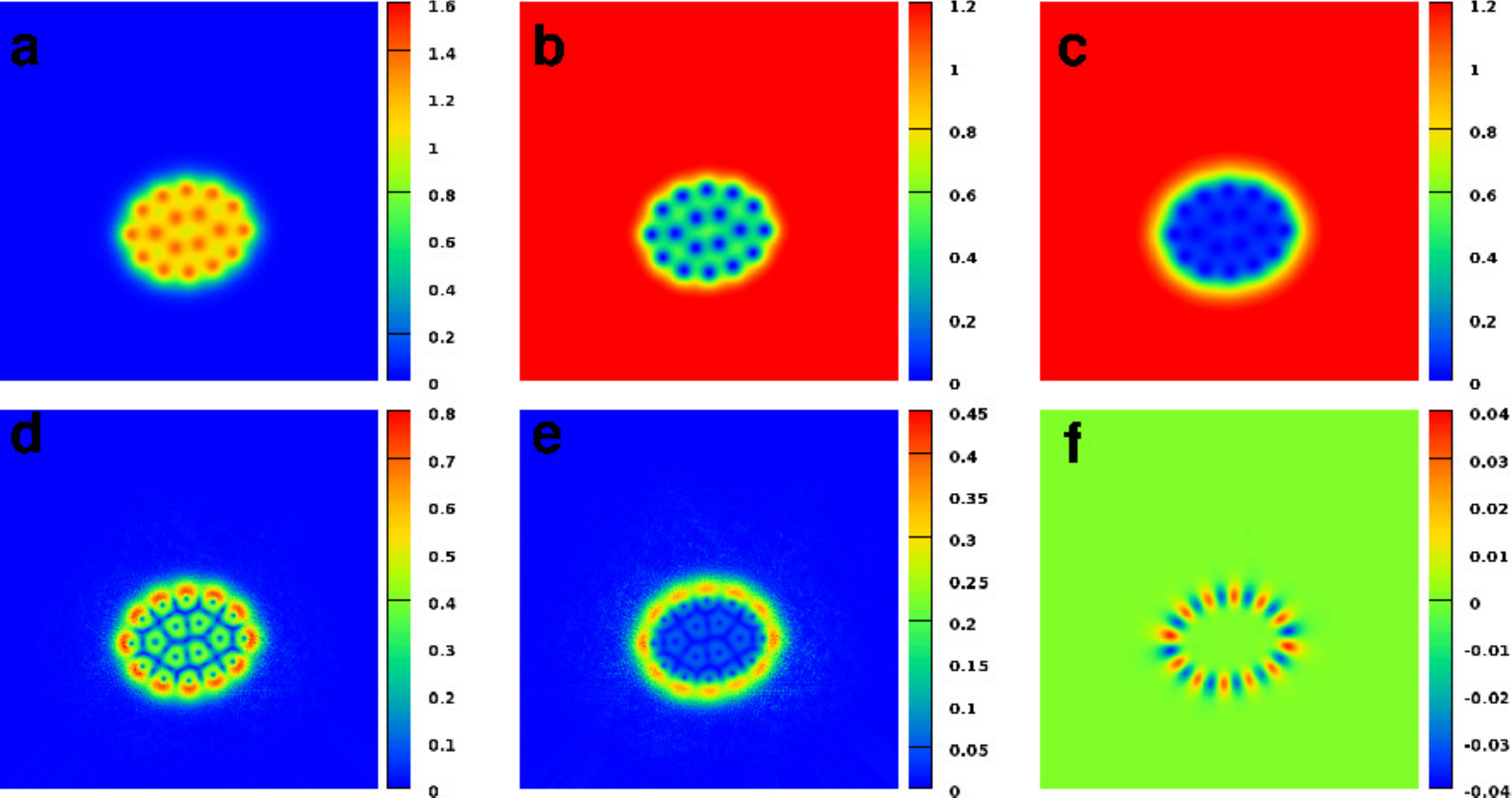}
 \hss}
\caption{
Elongated cluster of vortices in the same superconductor as in \Figref{fig-new2}, but with $N_v=18$ vortices. 
}
\label{fig-new4}
\end{figure*}

\begin{figure*}[!htb]
  \hbox to \linewidth{ \hss
 \includegraphics[width=1.0\linewidth]{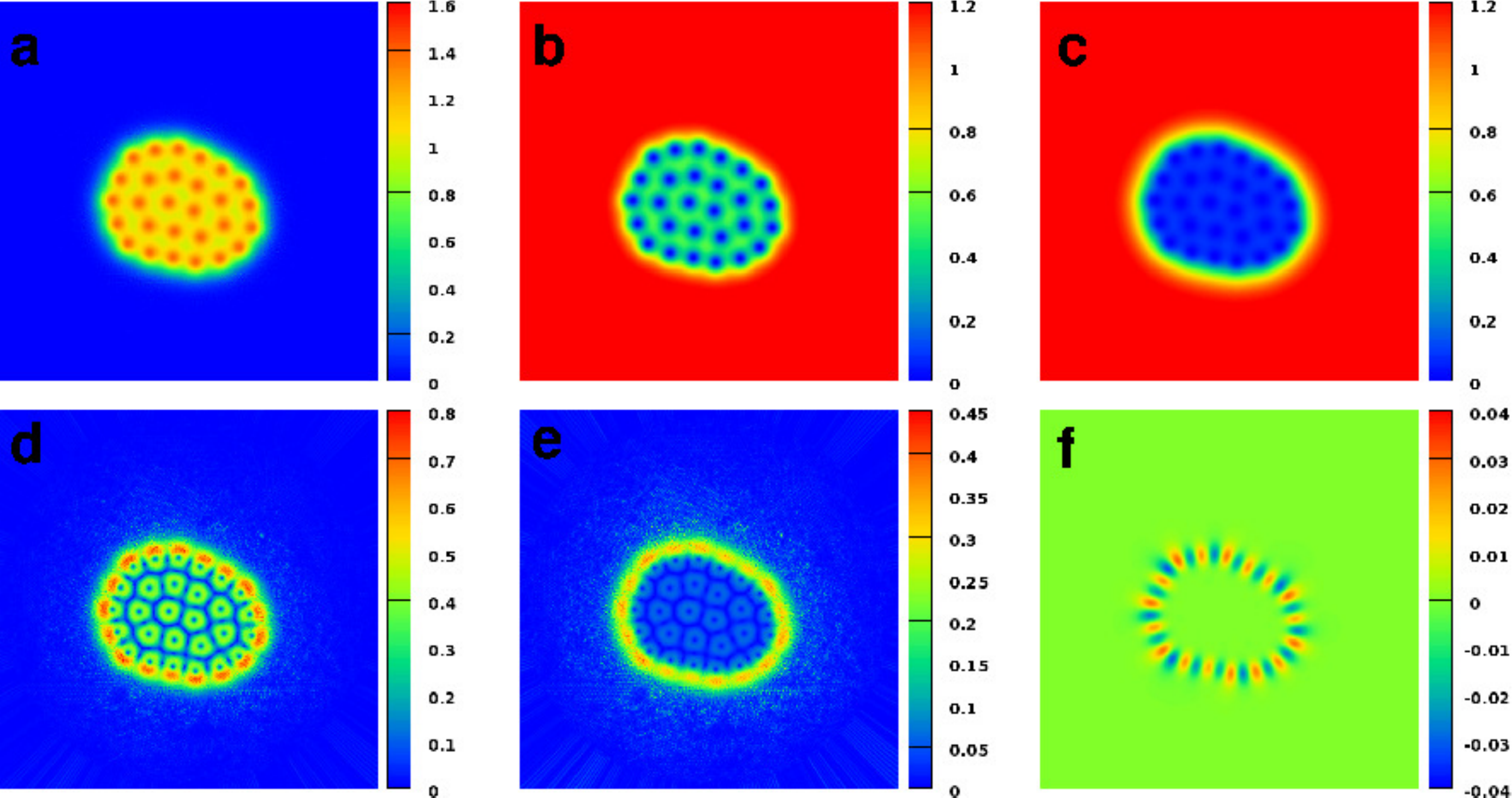}
 \hss}
\caption{
Bound state of $N_v=28$ vortices, for a superconductor with broken $U(1)\times U(1)$ symmetry, 
with the parameters as in \Figref{fig-new2}. 
This irregularly shaped cluster represents  a local minimum of the free energy.
The local minima originate
from the competing interactions yielding a 
complicated free energy landscape.
}
\label{fig-new5}
\end{figure*}

 The multibody forces in this nonlinear theory, originate in nontrivial  deformation
of vortices by their neighbors in a vortex cluster.
First the qualitatively new  physics which arises in these  clusters is the appearance of gradients 
of the phase difference $\nabla(\theta_1-\theta_2)$ between the condensate fields. It is 
clearly seen on the panels $\bf f$ from the plotted quantity $\Im(\psi_1^*\psi_2)$. One of 
the mechanisms of the generation of the phase difference which we observed was associated 
with  splitting of the vortex cores of the components $\psi_{1,2}$ driven by competing interactions. 
It leads to a ``dipole"-like configuration of the phase difference of the two 
components which in turn results in  contributions to muti-body forces
associated with the induced phase difference gradients. This splitting exists in cases of zero, { as well as finite  Josephson coupling}, 
though it is smaller in the later case. 
We observed also more complicated configurations like 
``quadrupoles" of the phase difference fields.
These configurations occurred when the vortices broke their
axial symmetry by relegating more
phase gradients in the areas where density 
was depleted by neighboring vortices. 
Another source of multibody forces was 
associated with nontrivial condensate densities modulations
for a group of several vortices.

Note that  the presence of 
gradients in the phase difference, along with gradients of the relative density of two condensates
is known to 
lead to contributions from self-generated Skyrme-like terms to magnetic energy density 
\cite{knots2}. This makes the physics of the two-component vortex cluster boundary 
and the resulting multibody forces, in general, a complicated nonlinear problem. 

With increased number of vortices the simulations frequently produced long-living 
bound states of irregular shapes like that shown on Figs. \ref{fig-new4} and \ref{fig-new5}.

Parameter sets where the system is close to type-I regime (like the one shown of Fig. (\ref{2AWJ-1}))
show that the transition (in the parameter space of the model) to type-I regime
manifests itself in a depletion of current densities  inside vortex clusters and relative
increase of current density on the cluster's boundary. In type-I regime the system
forms one circular cluster where the current is concentrated on the boundary (i.e. a single giant vortex).

 Note  that in single-component type-I superconductors the  domains of normal 
phase have a circular boundary only when the effects of stray fields outside sample are neglected. 
In finite samples of type-I superconductors the  stray  magnetic fields typically lead to formation 
of macroscopically large stripes of the normal phase rather than circular domains \cite{degenes}, 
though other geometries were also observed \cite{prozorov}. Similarly  in realistic experimental 
setups  especially for type-1.5 bilayers, stray magnetic fields could lead to vortex stripes rather 
than circular vortex clusters formation.

\begin{figure*}[!htb]
  \hbox to \linewidth{ \hss
 \includegraphics[width=1.0\linewidth]{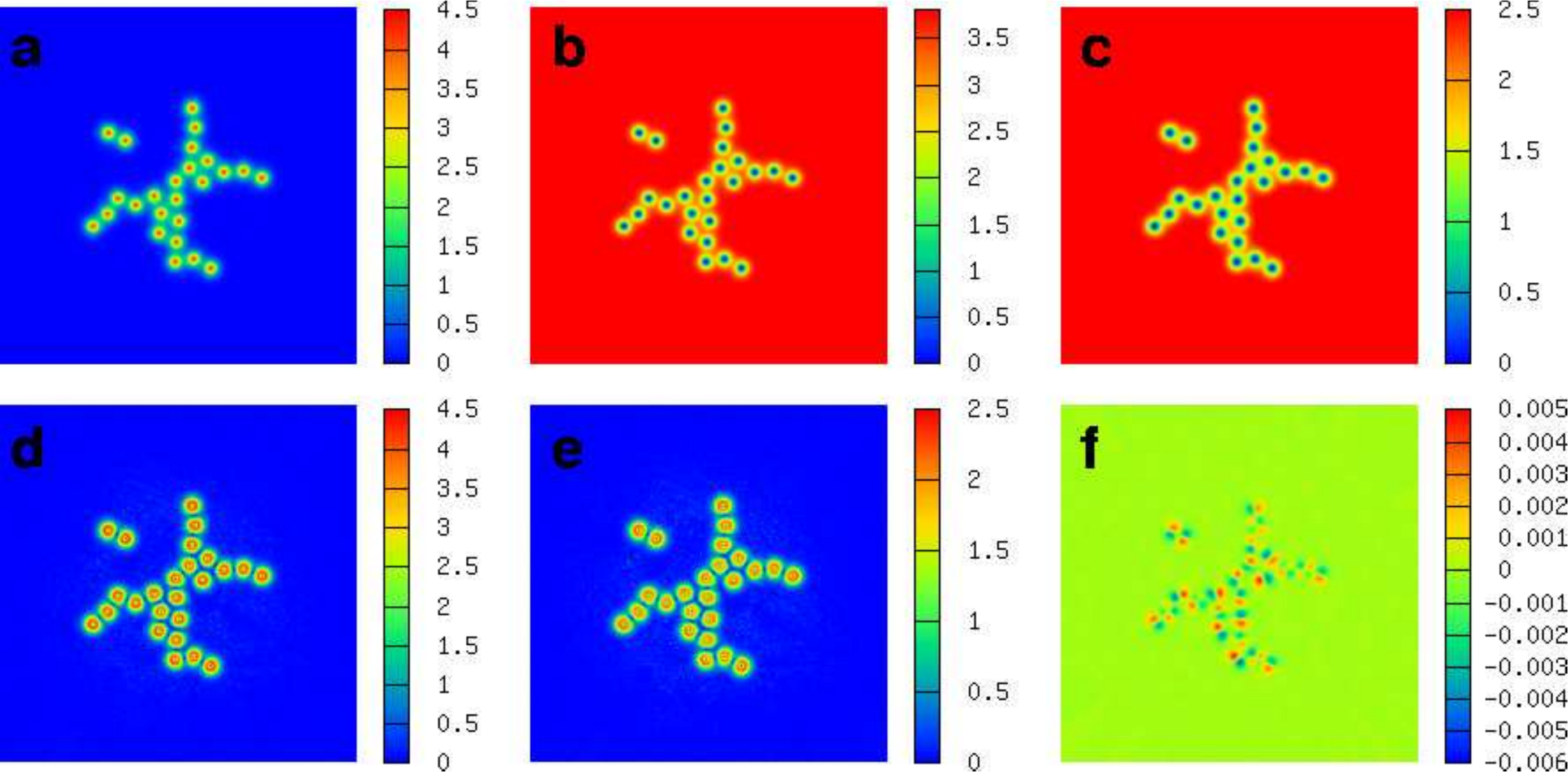}
 \hss}
\caption{
A  bound state of an $N_v=25$ vortex configuration in case when  
superconductivity in the second band is due to interband proximity effect 
and the  Josephson coupling is very strong $\eta=7.0$.
The initial configuration in this simulation was a giant vortex. Other parameters 
are $(\alpha_1,\beta_1)=(-1.00, 1.00)$, $(\alpha_2,\beta_2)=(3.00, 0.50)$,  $e=1.30$.
}
\label{1A1PSJ}
\end{figure*}

\section{Non-compact vortex clusters and non-pairwise interactions}
Next we study the structure formation in a regime with relatively strong
non-pairwise forces.
In this section we consider a situation where the  passive  (i.e. with positive $\alpha_2$) second band is coupled 
to the first band by extremely strong Josephson coupling $\eta=7.0$ (shown on \Figref{1A1PSJ}). This coupling 
imposes  a strong energy penalty both for disparities of the condensates variations and for phase 
difference. The two and three body forces in that regime are shown
in the middle column on \Figref{pan5}. Indeed the two-body forces in this case have
 only a weak non-monotonicity. Importantly, there is an anisotropy of three body forces shown in \Figref{pan5}.
 It clearly diminishes the energetic benefits of a triangle-like states compared to line-like vortex states. 
  Identifying the ground states in this regime  numerically (which involves four-body and higher order interactions)
 is much more complicated than in the 
previous cases. We get a flat and complicated energy landscape and the outcome of the 
 energy minimization strongly depends on initial configuration (see Appendix for
the description of numerical procedures). The 
\Figref{1A1PSJ} shows the typical non-universal outcome of the energy minimization in 
this case. Striking feature here is  formation of  vortex stripe-like configuration. 
Indeed it strongly contradicts the ground state structure  predicted by the two-body forces in this system.
Namely the axially symmetric two-body potentials with 
long range attraction and short-range repulsion (which we have in this case) do not allow 
stripe formation in the ground state configurations..
 Therefore the observation of the vortex stripes signals that the structure formation,
along with short-range character of attractive tail, is 
influenced by repulsive multi-body forces in these cases (in contrast to 
the structure formation in the previous section which was dominated by two-body forces). 
Note that even in this regime, the system 
exhibits self-induced gradients of the phase difference, in spite of the  strong Josephson 
coupling. In order to study the role  of initial conditions we consider the vortex ordering in 
a  similar regime but  starting with   30 vortices  placed at 
distances larger than the minimum of the two-body potential. If there were only two-body 
forces this vortex clusters would contract to minimize the energy. In contrast in the energy 
minimization process the vortex cluster first {  expands} (see the animation
of the evolution of the system in the numerical energy minimization  process in
the supplementary material \cite{julien}). 
Subsequently it breaks into a few sub-clusters and 
vortex chains. At the final stage 
of the evolution each cluster contracts. Final intervortex distances in 
each sub-cluster is smaller than intervortex distance in the initial state
shown on \Figref{1A1PSJb}.

Formation of highly disordered states and vortex chains due to
short-range nature of the attractive potentials and many-body forces was a generic outcome of the 
simulation in the similar type-1.5 regimes with strong Josephson coupling, in spite 
of negligible effects of ultra-fine numerical grid.

In this section we considered the regimes where the attractive part of
two-body interaction was relatively weak and of short range.
Physically, the fact that there are also multi-body forces 
which energetically penalize the hexagonal arrangement
in large groups of vortices  implies that at finite temperatures small clusters and 
irregular chain-like structures should easily form as well  for entropic and kinetic reasons.

\begin{figure*}[!htb]
  \hbox to \linewidth{ \hss
 \includegraphics[width=1.0\linewidth]{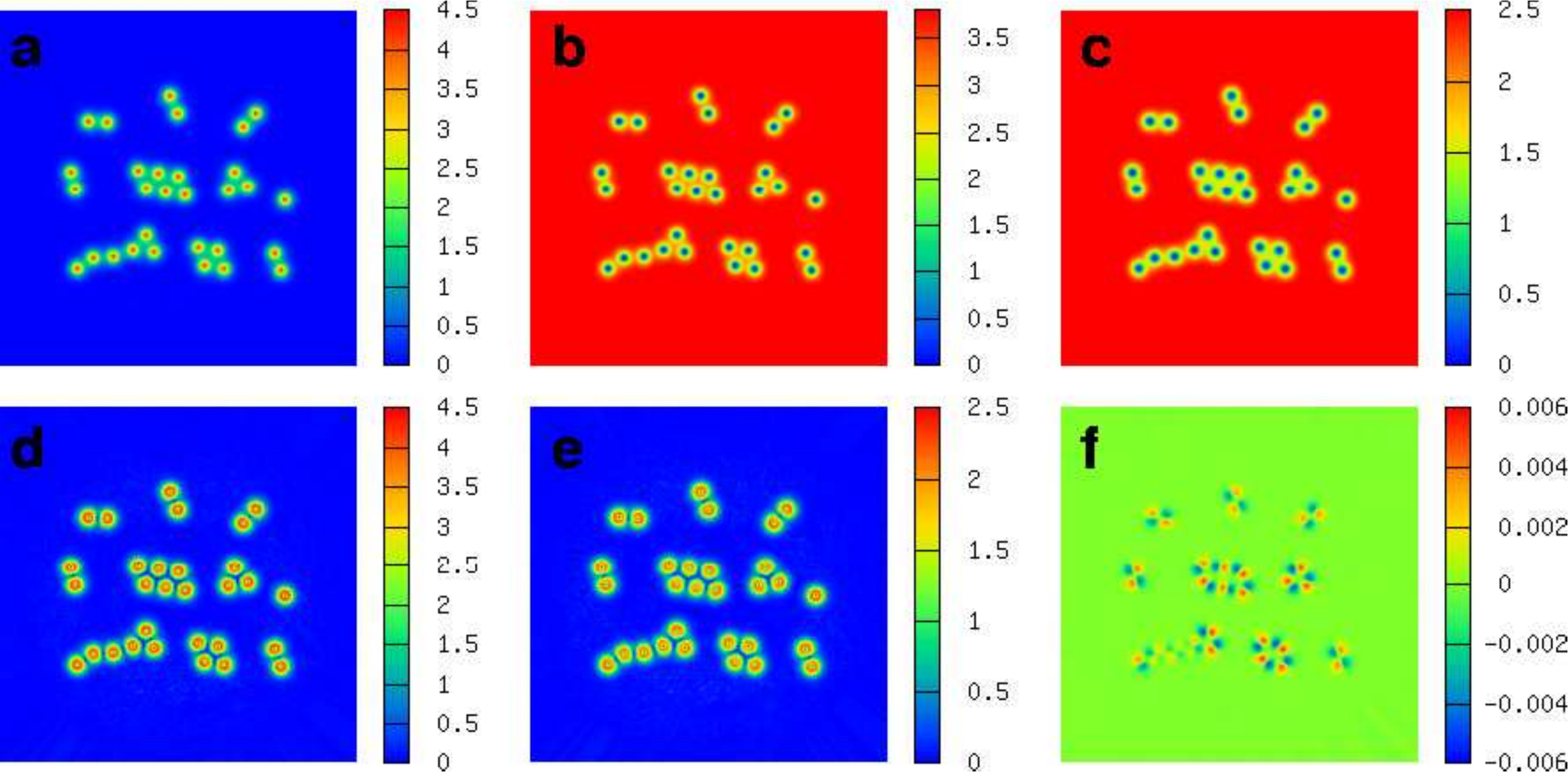}
 \hss}
\caption{
A  bound state of an $N_v=30$ vortex configuration 
for the system with parameters like in \Figref{1A1PSJ}
which was obtained using a dilute initial configuration
of vortices \cite{julien}.
}
\label{1A1PSJb}
\end{figure*}

\section{Conclusions.}
In conclusion, 
in this work we investigated structures of vortex clusters in the Semi-Meissner state and 
demonstrated  that {\it non-pairwise} interaction forces can be especially
important in multicomponent and layered superconductors. 
It results in the very rich physics associated with the vortex clusters in type-1.5 regime. Namely 
N-quanta clusters can be quite different from simple superpositions of N single 
vortex solutions. In general they should have multiple local minima in the energy
landscape. Thus  vortex clusters  can have extremely irregular shapes {\it even in absence of vortex pinning}.
The discussed above regimes can be directly probed experimentally in type-I/type-II
multilayers where intercomponent coupling can be tuned and be distinguished
from pinning effects by $IV$ characteristics.
The found here non-pairwise forces should similarly
be important in type-II multiband superconductors or layered structures
for understanding  compact configurations of pinned vortex clusters.

We also remark that in certain cases
the separation into vortex and Meissner domains
also implies phase separation into domains with different broken symmetries.
Consider $U(1)\times U(1)$ model. Even if the
second component there is not completely depleted in the vortex cluster, its density is
suppressed and as a consequence the magnetic  binding energy between
vortices with different phase windings ($\Delta \theta_1=2\pi,
\Delta \theta_2=0$) and  ($\Delta \theta_1=0, \Delta
\theta_2=2\pi$) can be arbitrarily small \cite{frac}. Moreover the vortex
ordering energies in the  component with more depleted density will
also be small. As a result, even small thermal fluctuation can drive
vortex sublattice melting transition \cite{Nature} in
a macroscopically large vortex droplet.  In that case the
fractional vortices in weaker component tear themselves off the
fractional vortices in strong component and form a disordered
state. Note that the vortex sublattice melting is associated with
the  phase transition from $U(1)\times U(1)$ to $U(1)$ state
\cite{Nature}. I.e. that vortex cluster where one sublattice has melted will represent
a domain of $U(1)$ phase (associated with the superconducting
state of strong component) immersed in domain of vortexless
$U(1)\times U(1)$ Meissner state. 
 
 We thank A. Gurevich and J.M. Speight for stimulating discussions.
The work is supported by the NSF CAREER Award No. DMR-0955902, by  
Swedish Research Council, and by the Knut and Alice Wallenberg
Foundation through the Royal Swedish Academy of Sciences fellowship.  
Computations were performed at the National Supercomputer Center (NSC) in Linkoping, Sweden.

\appendix

\section{Finite difference energy minimization} \label{Numerics-fd}
To calculate intervortex interaction energies we use finite difference energy minimization.
%
Ground states of vortex systems and inter-vortex interaction energies are found by minimizing this functional subject to relevant constraints, such as vortex positions. 
To do this numerically, we discretize the system on a regular grid. To have the 
numerical results unbiased we use a non-adaptive grid where the grid spacing $h$ is the same everywhere in the domain.  
The Hamiltonian is then discretized using the  finite difference approach:

Gradients are defined as 
\bea
(\nabla f)_{i,i+1} = \frac{f(i+1)-f(i)}{h}
\eea
and magnetic flux is computed by line integration

\bea
B_{i,i+1,j,j+1}=\frac{1}{h^2} \oint_{\omega}  \bar{A}\cdot d\bar{r}
\eea 
where $\omega$ is the square with the corners $i,i+1,j,j+1$. 
The energy density in the grid point $(i,j)$ then depends on function values in $i,j$ and its neighbors. 

The optimization scheme which is used in the first part of the paper
is a modified version of the Newton-Raphson method. 
To minimize boundary effects we use free boundary conditions.
Vortices are inserted using various initial configurations.

In order to calculate the inter-vortex interaction energy, we have to fix the position of vortices. 
Fixing a vortex position requires a special care to avoid the situation where  the pinning 
 substantially affects the vortex solution.
We fix the vortex position by the following method.
In the vortex center the condensate density is zero. We then fix the density only of the central point of the
dominant component $|\psi_i|$ of the vortex to be
 zero in a given position of numerical grid. This effectively prevents the vortex 
from moving but does not prevent core splitting of  $|\psi_1|$ and $|\psi_2|$ due to competing interactions. This
``point pinning" method also has advantage of 
being a ``minimially invasive"  since only the position of  the core singularity
is fixed. Thus it allows calculate medium- and long-range forces
with greatest accuracy. 

However, at the same time, obviously, this method does not work for too short
intervortex separation.
For too short vortex separation it leads to the following 
easily identifiable artifact:
a vortex core of one of the vortices elongates to be zero
at both pinning centers allowing  the second vortex to unpin and escape, while
satisfying the energy minimization constraint. 
This behavior can be easily remedied by different pinning schemes.
However for consistency and also  because very short-distance intervortex forces are irrelevant for the
questions studied in this paper  we use one pinning procedure.

Convergence is determined as follows: 

1) We choose a particular grid spacing $h_1$ and number of grid points $N_1=N_{1x}\cdot N_{1y}$ giving a system size $L_x=h\cdot (N_{1x}-1),\;L_y=h\cdot (N_{1y}-1)$. 
Then we minimize the energy until it does not change in a few thousand iterations. This gives us $E(h_1)$.

2) We decrease grid spacing $h$ by a factor of 2 or 3 while retaining the system size $L_x,L_y$ using spline interpolation. Then, we once again iterate until the energy does not change 
in a few thousand iterations, giving us $E(h_2)$ and so forth. We then determine convergence from 

\bea
\frac{E(h_n)-E(h_{n+1})}{E(h_n)} = C.
\eea
We use grid sizes up to $N\approx10^7$ which gives very high accuracy, typically $C<10^{-4}$.

\section{Finite element energy minimization}\label{Numerics-fem}

In the second part of the paper we use the uncontrained energy minimization.
Bound state vortex configurations are minima of Ginzburg-Landau energy \Eqref{ind_energy}. 
This means that functional minimization of \Eqref{ind_energy}, from an appropriate initial 
state describing several flux quanta, should lead to bound state (if it exists). 
We consider the two-dimensional problem $\mathcal{F}=\int_\Omega F$ defined on 
the bounded domain $\Omega\subset\mathbbm{R}^2$, supplemented by `open' boundary conditions.

The variational problem is defined for numerical purpose using a finite element formulation
provided by the Freefem++ \cite{Freefem} framework. Discretization within finite element 
formulation is done via a (homogeneous) triangulation on $\Omega$, based on Delaunay-Voronoi algorithm.
Functions are decomposed on a continuous piecewise quadratic basis on each triangle.

Contrary to the numerical method used in the first part, the accuracy does not depend only on the `number of grid points'.
The accuracy of such method is controlled through the number of triangles,  (we typically used  $~10^5$), the order of expansion  of the basis on each triangle (P2 elements are 2nd order polynomial basis on each triangle), and also the order of the quadrature formula 
for the integral on the triangles. 

Once the problem is mathematically well posed, a numerical optimization algorithm is used to solve the variational nonlinear problem
(i.e. to find the minima of $\mathcal{F}$). We used here  {Nonlinear Conjugate Gradient} method
Algorithm is iterated until relative variation of the norm of the gradient of the functional  $\mathcal{F}$ with respect to all degrees of 
freedom is less than $10^{-6}$. To be sure that our results are not numerical artifacts of this particular minimization scheme, 
we also performed standard Steepest Descent calculations and checked it leads to similar results.

Minimization starts with an initial guess: a field configuration carrying the $N_v$ flux quanta described by
\Align{Initial_Guess1}{
\Phi_a &= u_a\prod_{i=1}^{N_v} 
\sqrt{\frac{1}{2} \left( 1+\tanh\left(\frac{4}{\xi_a}({\cal R}_i(x,y)-\xi_a) \right)\right)}
\mathrm{e}^{ i\Theta_i}\, ,~~  \nonumber \\
\vec{A}&=
\frac{1}{e{\cal R}}\left(\sin\Theta,-\cos\Theta \right)\,,
}
where $a=1,2\,$, $u_a\,$ is the vacuum expectation value of each scalar field, 
the parameter $\xi_a$ give the core sizes while $\Theta\,$ and $\cal{R}\,$ are 
\Align{Initial_Guess2}{
\Theta(x,y)&=\sum_{i=1}^{N_v}\Theta_i(x,y) \,,  \nonumber\\
\Theta_i(x,y)&=\tan^{-1}\left(\frac{y-y_i}{x-x_i}\right)\,, \nonumber\\
{\cal R}(x,y)&=\sum_{i=1}^{N_v}{\cal R}_i(x,y)\,, \nonumber\\
{\cal R}_i(x,y)&=\sqrt{(x-x_i)^2+(y-y_i)^2}\,. 
}
$(x_i,y_i)$ are the initial position of a given vortex. 
 Then, all degrees of freedom are relaxed 
simultaneously without \emph{any} constraint to obtain high accurate solutions of the Ginzburg-Landau equations.

The initial guess \Eqref{Initial_Guess1}
allows starting from various very different initial configurations, depending on the values of the 
$(x_i,y_i)$. Since we know from two-body calculations, the preferred distance between two vortices, 
we choose to start either in the \emph{repulsive} or in the \emph{attractive} tail of the two-body interaction potential.
Animations showing the evolution of the system from these various initial configurations, during
the described above energy minimization is available as online
supplementary material  \cite{julien}. Note that we do not solve a dynamical problem here, the main
purpose of these movies is that they reflect the structure of the free energy landscape and to give
intuitive illustration of intervortex forces. 
For a reader interested in the a relationship between this energy minimization 
dynamics and the dynamics of Time Dependent Ginzburg-Landau model 
we refer to Ref. \cite{du}.

The method described in the first part of the paper was also used for unconstrained simulations to
doublecheck some of the results.

\end{document}